\documentclass{jaa}
\usepackage{graphicx}
\usepackage{caption}
\usepackage[font=scriptsize]{caption}
\usepackage{subcaption}
\usepackage{amssymb}
\usepackage{geometry}
\usepackage{multirow}
\usepackage{natbib}
\usepackage{array}
\usepackage{float}
\usepackage{url}
\usepackage{threeparttable}

\begin{document}\sloppy

\title{PAH and nbL Features Detection in Planetary Nebulae NGC 7027 and BD $+$30$^\circ$~3639 with TIRCAM2 Instrument on 3.6m DOT}
\author{RAHUL KUMAR ANAND\textsuperscript{1}, SHANTANU RASTOGI\textsuperscript{*1}, BRIJESH KUMAR\textsuperscript{2}, ARPAN GHOSH\textsuperscript{2}, SAURABH SHARMA\textsuperscript{2}, D. K. OJHA\textsuperscript{3} and S. K. GHOSH\textsuperscript{3}}
\affilOne{\textsuperscript{1}Department of Physics, DDU Gorakhpur University, Gorakhpur 273009, India.\\}
\affilTwo{\textsuperscript{2}Aryabhatta Research Institute of Observational Sciences, Manora Peak, Nainital-263001, India.\\}
\affilThree{\textsuperscript{3}Tata Institute of Fundamental Research, Homi Bhabha Road, Colaba, Mumbai 400 005, India.}


\twocolumn[{
\maketitle

\corres{shantanu\_r@hotmail.com}


\begin{abstract}
High resolution infrared imaging observations of the young Planetary Nebulae NGC 7027 and BD +30$^\circ$~3639, taken with the newly installed TIFR Infrared Camera-II (TIRCAM2) on 3.6m Devasthal Optical Telescope (DOT), ARIES, Nainital, are being reported. The images are acquired in J, H, K, polycyclic aromatic hydrocarbon (PAH) and narrow-band L (nbL) filters. The observations show emission from warm dust and PAHs in the circumstellar shells. The imaging of the two objects are among the first observations in PAH and nbL bands using TIRCAM2 on DOT. The NGC 7027 images in all bands show similar elliptical morphology with $\sim$6$^{\prime\prime}\hspace{-0.13cm}$.7 and $\sim$4$^{\prime\prime}\hspace{-0.13cm}$.5 semi-major and semi-minor axes. Considering size up to 10\% of peak value the nebula extends upto 8$^{\prime\prime}$ from the central star revealing a multipolar evolution. The relatively cooler BD +30$^\circ$~3639 shows a rectangular-ring shaped nebula. In J and H bands it shows an angular diameter of $\sim$8$^{\prime\prime}$, while a smaller $\sim$6$^{\prime\prime}\hspace{-0.13cm}$.9 size is observed in K, PAH and nbL bands. The 3.28~$\mu$m emission indicates presence of PAHs at about 6000 and 5000~AU from the central stars in NGC 7027 and BD +30$^\circ$~3639 respectively. Analysis suggests domination of neutral PAHs in BD $+$30$^\circ$~3639, while in NGC 7027 there is higher ionization and more processed PAH population.

\end{abstract}


}]


\doinum{12.3456/s78910-011-012-3}
\artcitid{\#\#\#\#}
\volnum{000}
\year{0000}
\pgrange{1--}
\setcounter{page}{1}
\lp{1}

\section{Introduction}

Planetary Nebulae (PNe) can be easily identified from other celestial objects for their strong emission lines in optical and infrared spectral regions. A star that has initial mass less than 8 \(M_\odot\) ends its life through a period of extreme mass loss. As the star moves up the asymptotic giant branch (AGB), it undergoes severe mass-loss ($10^{-7}$--$10^{-4}$~\(M_\odot\)~per~year). The post-AGB or proto Planetary Nebulae phase is a short rapidly varying phase evolving towards the PNe phase (\citet{Winckel2003, Kwok2007} and references therein). During this phase the circumstellar dust shell moves away from the central star and temperature decreases from $\sim$400 to 100~K \citep{Bedijn1987, Zijlstra1992}. As the circumstellar shell moves away enough to expose the central hot core, the PNe is revealed. The central hot core ionizes the circumstellar shell exhibiting emission lines and continuum. Most PNe deviate from spherical symmetry and show spectacular bipolar or multipolar structures. The short transition period from post-AGB to evolved PNe is not well understood. This is an important evolutionary phase as the circumstellar shells also provide environments for formation of molecules and chemical evolution \citep{Ziurys2006, Agundez2008, Sande2019}. The post-AGB, proto-PNe with carbon rich circumstellar shells are also considered to be the breeding ground for polycyclic aromatic hydrocarbon (PAH) molecules \citep{Frenklach1989, Cherchneff1992, Tielens2008} seen through their mid-infrared emission features.

Near- and mid-IR observations provide the high spatial and spectral resolution required to study and understand the form and content of PNe. The observed flux is a function of several factors viz. hydrogen bound-free continuum emission, recombination and collisionally excited lines, molecular hydrogen emission and continuum components from hot dust of the medium. There are also ionic emission lines (e.g. Ne, Ar, S, Mg etc.), broad silicate feature near 9.7 $\mu$m, and infrared emission band features. Many bright infrared nebulae show mid-IR emission features attributed to PAH molecules pumped by background UV radiations and are referred to as Aromatic Infrared Bands (AIBs) \citep{Leger1984, Allamandola1985, Tielens2008, Rastogi2013}. Several PNe have been observed from ground in the spectral range 8-13 $\mu$m \citep{Russell1977, Aitken1982, Roche1986}, between 7.5 and 23 $\mu$m with IRAS low resolution spectrometer \citep{Pottasch1986}, Kuiper Airborne Observatory \citep{Cohen1989}, Infrared Space Observatory (3.6-160 $\mu$m) \citep{Persi1999}, and Stratospheric Observatory for Infrared Astronomy (SOFIA) (0.3-1600 $\mu$m) \citep{Reinacher2018}. All observations point towards the complex physical and chemical evolution in these objects.

There is correlation between the presence of molecular emission and the morphology of PNe, such that objects rich in molecular material are bipolar or butterfly nebulae \citep{Hora1999, Kastner1996, Huggins1996, Huggins1989, Zuckerman1988}. High progenitor mass also correlates with this bipolar morphology type \citep{Corradi1995}. This suggests that the higher mass AGB progenitors that have high mass loss rates produce dense, long lived molecular envelopes \citep{Hora1999}. The carbon rich envelopes are considered primary source of interstellar PAHs \citep{Latter1991}, which seed the growth of amorphous carbon dust \citep{Tielens2008}. Yet, the PAH emission features are rarely observed in AGB stars, probably due to the effective low temperature of AGB photosphere which are unable to excite the PAH vibrational modes. The PAH features are observed in spectra of post-AGB and PNe where the circumstellar dust shell is heated by the hard UV radiation field from the hot core. Thus, indication of the location of photo-excited PAHs can be useful in the understanding of the molecular and morphological evolution.

The observations in near-IR and PAH emission are difficult from ground-based telescopes. The newly installed TIFR Near Infrared Imaging Camera-II (TIRCAM2), attached to the main axial port of 3.6 m Devasthal Optical Telescope (DOT), ARIES, Nainital \citep{Baug2018, Kumar2018, Sagar2019}, provides an unique opportunity to study the 3.28~$\mu$m PAH feature from ground. In the present communication the TIRCAM2 is used for near-IR, conventional as well as narrow wavebands, imaging studies of the two young and carbon rich planetary nebulae NGC 7027 and BD +30$^\circ$~3639. NGC 7027 is in a rapid and key moment in its evolution when it is transiting from predominantly neutral molecular envelope to an ionized one \citep{Kwok2000}. BD +30$^\circ$~3639 is a young planetary nebula that hosts Wolf-Rayet [WC] Central star and is round, highly symmetric low-excitation nebula \citep{Balick1987}. Both objects have significant AIB features \citep{Peeters2002, Diedenhoven2004}. The observations in PAH band give insight into the location and environment where PAHs have abundance.

\section{Observations and Data Reduction }

\subsection{TIRCAM2 Imaging Observation}

\begin{figure}
    \includegraphics[height=12cm, width=\columnwidth]{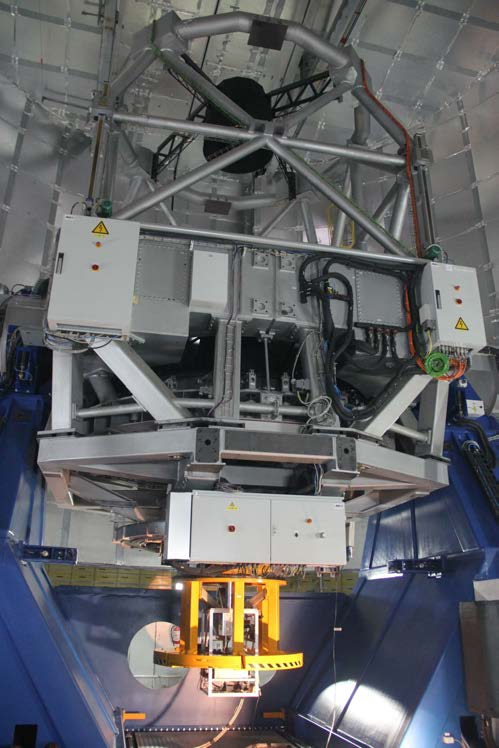}
    \caption{The TIRCAM2 mounted at the main axial port of the 3.6 m DOT.}
    \label{fig1}
  \end{figure}

The near-IR imaging of the two objects NGC 7027 and BD +30$^\circ$~3639 are performed in J, H, K, PAH and nbL (narrow-band L) filters, with wavelengths centring at 1.20, 1.65, 2.19, 3.28 and 3.59 $\mu$m, respectively. Among the observed AIBs a significant feature attributed to aromatic C-H stretch vibrations falls around 3.28 $\mu$m, therefore the corresponding filter is called PAH filter. The TIRCAM2 is attached to the main axial port of Cassegrain focus (Figure~\ref{fig1}) of the 3.6 m DOT (latitude 29$^\circ$.1971 N, longitude 79$^\circ$.6841 E, altitude: 2450 m) and provides sub-arcsecond spatial resolution. Both the targets are good sources for ground based observation due to their compactness and high surface brightness in the infrared \citep{Werner2014}. These are among the first observations in PAH band at 3.28 $\mu$m ($\bigtriangleup\lambda$ $\sim$0.06 $\mu$m) using TIRCAM2 on DOT. The observations of NGC 7027 and BD +30$^\circ$~3639 are recorded on the nights of 2017 October 13 and 2018 May 10 respectively.

The TIRCAM2 is a closed cycle Helium cooled near-IR imaging camera equipped with a larger Raytheon 512 x 512 pixels InSb Aladdin III Quadrant focal plane array (FPA). The operating temperature of detector is 35 K. The InSb array in TIRCAM2 is sensitive to photons in the 1-5 $\mu$m wavelength bands. However, the optics of TIRCAM2 restricts the operating wavelengths to below $\sim$3.8 $\mu$m. The camera has been designed to accommodate seven selectable standard near-IR filters \citep{Naik2012, Baug2018}, which can be used for imaging observations. Field of view (FoV) of the TIRCAM2 on DOT is $\sim$86$^{\prime\prime}\hspace{-0.13cm}$.5 x 86$^{\prime\prime}\hspace{-0.13cm}$.5 with a pixel scale of 0$^{\prime\prime}\hspace{-0.13cm}$.169 $\pm$ 0$^{\prime\prime}\hspace{-0.13cm}$.002. At Devasthal, the full width at half maxima (FWHM) of point source PSF (point spread function) during the observation nights was typically sub-arcsecond in all JHK bands.

During the observations standard strategy was adopted, which involves acquisition of dark frames and observations of flat frames with each filter during both morning and evening twilights. Between three-hundred to four-hundred different raw images were taken using JHK, PAH and nbL filters (Table~\ref{tbl-2}). The source position was dithered to different positions on the array, ensuring that good pixels are present in all location of the image. On 2017 October 13 the K-band PSF was $\sim$0$^{\prime\prime}\hspace{-0.13cm}$.76 and on 2018 May 10 it was $\sim$0$^{\prime\prime}\hspace{-0.13cm}$.62. For NGC 7027, four frames of 20s exposure have been observed in J and H bands while in the K-band seven frames of 12s exposure were acquired to avoid detector saturation. In a similar way three frames of 8s exposure have been observed for BD $+$30$^\circ$~3639 in J and H bands and for the K band five frames of 5s exposure time were acquired. 

Due to large background thermal emission in infrared, several hundred frames are taken in PAH and nbL bands with short exposure of 50ms for each frame in each dithered position for both the targets. Observation of nearby standard stars was also done, immediately before or after the observation of target source for photometric calibration. Dark frames were taken into account for the excess counts due to dark current in the detector. Flat-fielding is done using flat frames to remove non-uniformities in the science frames due to variation in pixel to pixel sensitivity and possible distortions caused by optics.
 
\subsection{\textbf{Data Reduction}}

The Image Reduction and Analysis Facility (IRAF) software\footnote{http://iraf.net/} is used for photometric reduction. Standard procedure is followed, first each scientific frame was subtracted with corresponding exposure time master dark frame, which is median combine of all respective dark frames. Then the dark corrected frames were passed through flat fielding correction, i.e. divided by normalized master flat frame of the same filter to calibrate for any pixel to pixel sensitivity variation. The near-IR night sky brightness is influenced by OH emission lines, water vapour, zodiacal emission, thermal emission from the atmosphere and the telescope, and moonlight \citep{Content1996, Sanchez2008}. This make the near-IR sky few hundred times brighter than the sky in optical-bands and its removal from near-IR images of target source is one of the key steps in the data reduction process. In this work, the source position was dithered to different positions on the array, making it possible for the images of target itself to be used for sky subtraction. Since the observed target field was not crowded, therefore it was possible to construct master sky frame by median combining the science target frames for corresponding near-IR-band. Finally, the resultant same filter dithered science frames were aligned and added together, yielding a good signal to noise ratio image of the bright source.

The DAOPHOT-II \citep{Stetson1992} package was used for photometric calibration of target sources. Both the PN objects are isolated in the science frames therefore aperture photometry is carried out in the J, H, K bands. The instrumental magnitude of the target sources was calibrated with respect to JHK standard sources \citep{Hunt1998} available on the two nights. The offset value of magnitude of standard source in JHK band, has been directly used to calculate the magnitude of both the target PNe. The photometric results are listed in Table~\ref{tbl-3}. For the photometry of NGC 7027 in K-band DAOPHOT-II gave error, probably because a large part of the K-band image overlaps bad patches in the lower side of the array. Measurements of various parameters viz. angular size of the objects, contour plots etc. are performed on SAO DS9 \citep{saods9}, which is a powerful tool to visualize astronomical data and also for its analysis.

\section{\textbf{Results}}

\subsection{\textbf{NGC 7027}}

\begin{figure*}
     \centering
       \includegraphics[width=\textwidth]{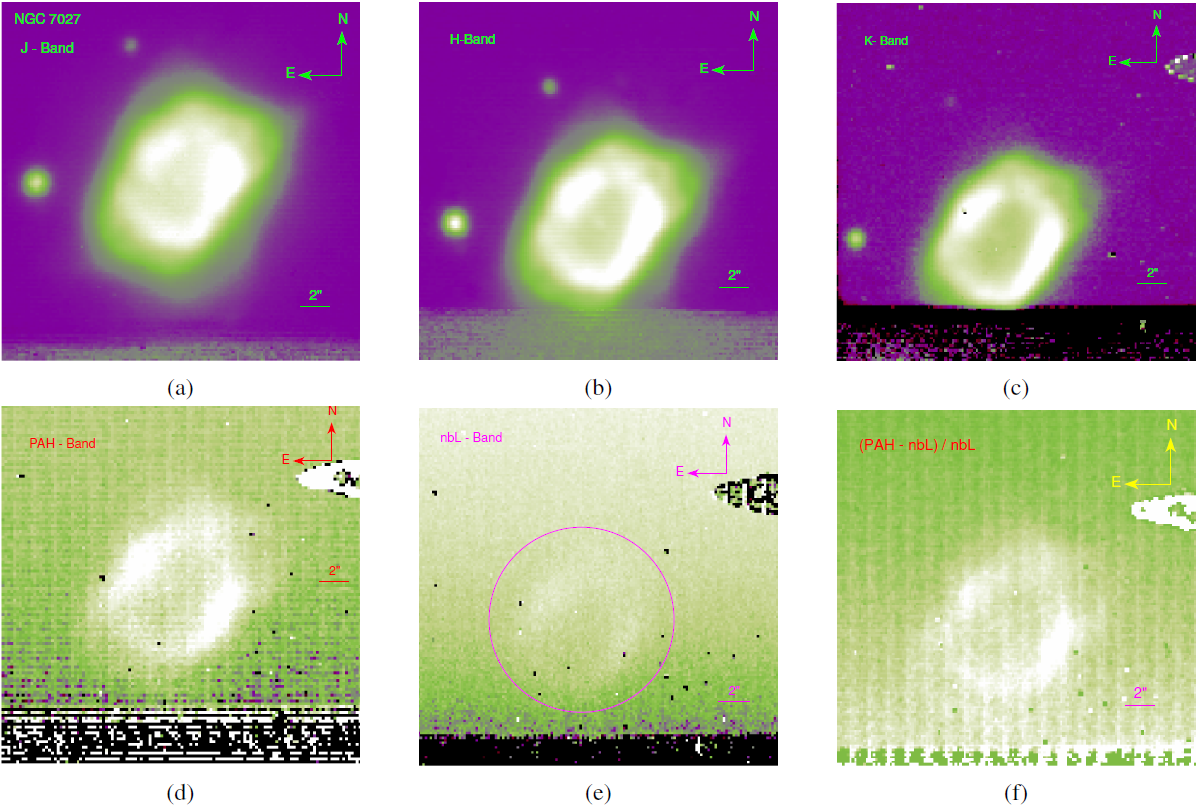}
        \caption{High-resolution images of NGC 7027 taken with TIRCAM2 on 3.6m DOT: (a) J, 1.2 $\mu$m; (b) H, 1.65 $\mu$m (c) K, 2.19 $\mu$m; (d) PAH, 3.28 $\mu$m; (e) nbL, 3.59 $\mu$m; (f) normalized continuum difference image, $(I_{PAH} - I_{nbL})/I_{nbL}$.}
        \label{fig2}
\end{figure*}

The final processed images in the five bands are shown in Figure~\ref{fig2}. The nbL-band ($\lambda$\textsubscript{cen} $\sim$3.59 $\mu$m, $\bigtriangleup\lambda$ $\sim$0.07 $\mu$m) falls close to the PAH-band ($\lambda$\textsubscript{cen} $\sim$3.28 $\mu$m, $\bigtriangleup\lambda$ $\sim$0.06 $\mu$m), but will have no contribution of the emission due to the C-H stretch mode AIB. Therefore, the nbL observation may be considered a continuum feature in the PAH-band. A normalized continuum difference image, $(I_{PAH} - I_{nbL})/I_{nbL}$, highlighting the emission from the AIB$_{3.28}$ is shown in Figure~\ref{fig2}(f). The central star of the NGC 7027 can be deciphered in the J and H bands while in the K band image it is too faint. It is also not visible in the PAH and nbL bands, implying that there is negligible near-IR emission from the central region. Two bright regions, seen in all the images along NE- and NW-direction, correspond to a surrounding torus of material \citep{Meixner1993}. These two bright regions have angular thickness $\sim$1$^{\prime\prime}\hspace{-0.13cm}$.91 and $\sim$2$^{\prime\prime}\hspace{-0.13cm}$.75 respectively and are symmetrically placed with respect to a NW - SE axis. The apparent mean angular diameter of NGC 7027 is measured as $\sim$14$^{\prime\prime}$ (Table~\ref{tbl-4}). The distance from the central star shows that major emissions from the nebula of NGC 7027 occur $\sim$2$^{\prime\prime}\hspace{-0.13cm}$.51 away.

\begin{figure*}
     \centering
        \includegraphics[width=\textwidth]{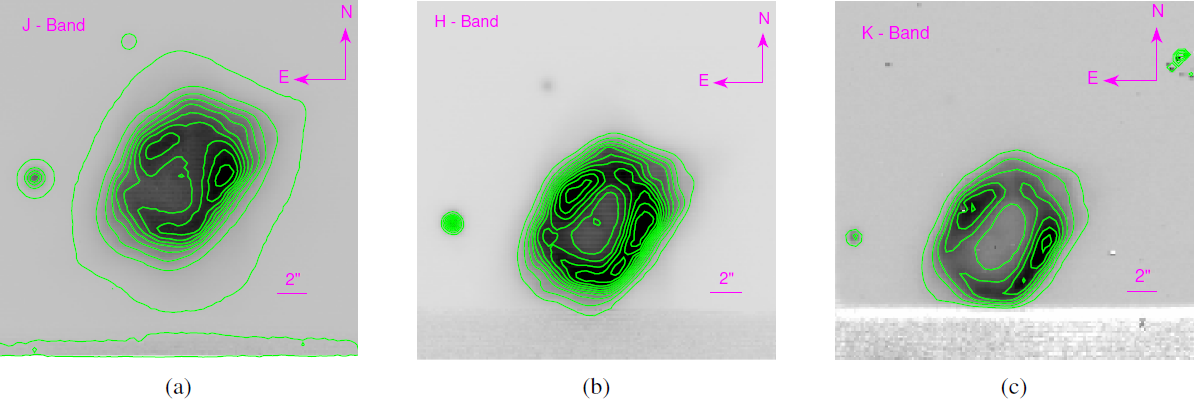}
        \caption{Contour images of NGC 7027 at: (a) J, 1.2 $\mu$m; (b) H, 1.65 $\mu$m; (c) K, 2.19 $\mu$m. Contours are separated by (a) 10\%, (b) 7.3\% and (c) 4\% of corresponding peak values.}
        \label{fig3}
\end{figure*}

The images of NGC 7027 (Figure~\ref{fig2}) demonstrate a nearly identical elliptical morphology with semi-major and semi-minor axes of $\sim$6$^{\prime\prime}\hspace{-0.13cm}$.7 and $\sim$4$^{\prime\prime}\hspace{-0.13cm}$.5, respectively. The major axis being inclined at an angle $\sim$-30$^\circ$ from North, which is consistent with the morphology of the H~II and photo-dissociation regions \citep{Latter2000, Cox2002}. In each band there are faint extended features that are more rectangular than elliptical. This deviation is better visualized in contour J, H, K images shown in Figure~\ref{fig3}. Considering pixel values of upto 10\% of the measured peak pixel value, the asymmetric features extend up to $\sim$8$^{\prime\prime}$ from the central star in the $\sim$-55$^\circ$ NW - SE direction. This is similar to the Paschen-$\alpha$ result of \citet{Lau2016}.

\begin{figure}
    \includegraphics[height=8cm, width=\columnwidth]{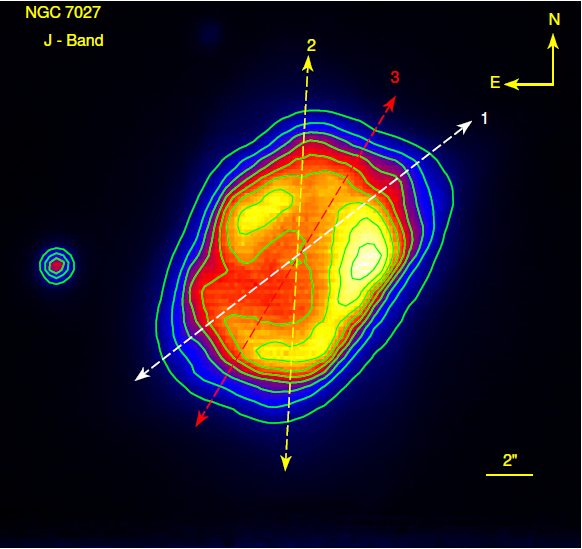}
    \caption{False color image of contour plot of NGC 7027 in J-Band with 1,2 and 3 outflow directions. The bi-directional arrows labelled 1-3 correspond to the direction of the outflows identified by \citet{Cox2002}. Outflow 1 (white) is believed to be the most recent and/or powerful outflow.}
    \label{fig4}
  \end{figure}

The J band contour map, Figure~\ref{fig4}, indicates outflows along the bi-directional arrows 1, 2, and 3. These correspond to the outflows identified by \citet{Cox2002} at $\sim$-55$^\circ$, $\sim$4$^\circ$ and $\sim$-28$^\circ$. The PAH and nbL features resemble elliptical morphology with a position angle along arrow 3 ($\sim$-28$^\circ$), while the extended features incline along arrow 1 ($\sim$-55$^\circ$).

\subsection{\textbf{BD $+$30$^\circ$~3639}}

\begin{figure*}
     \centering
         \includegraphics[width=\textwidth]{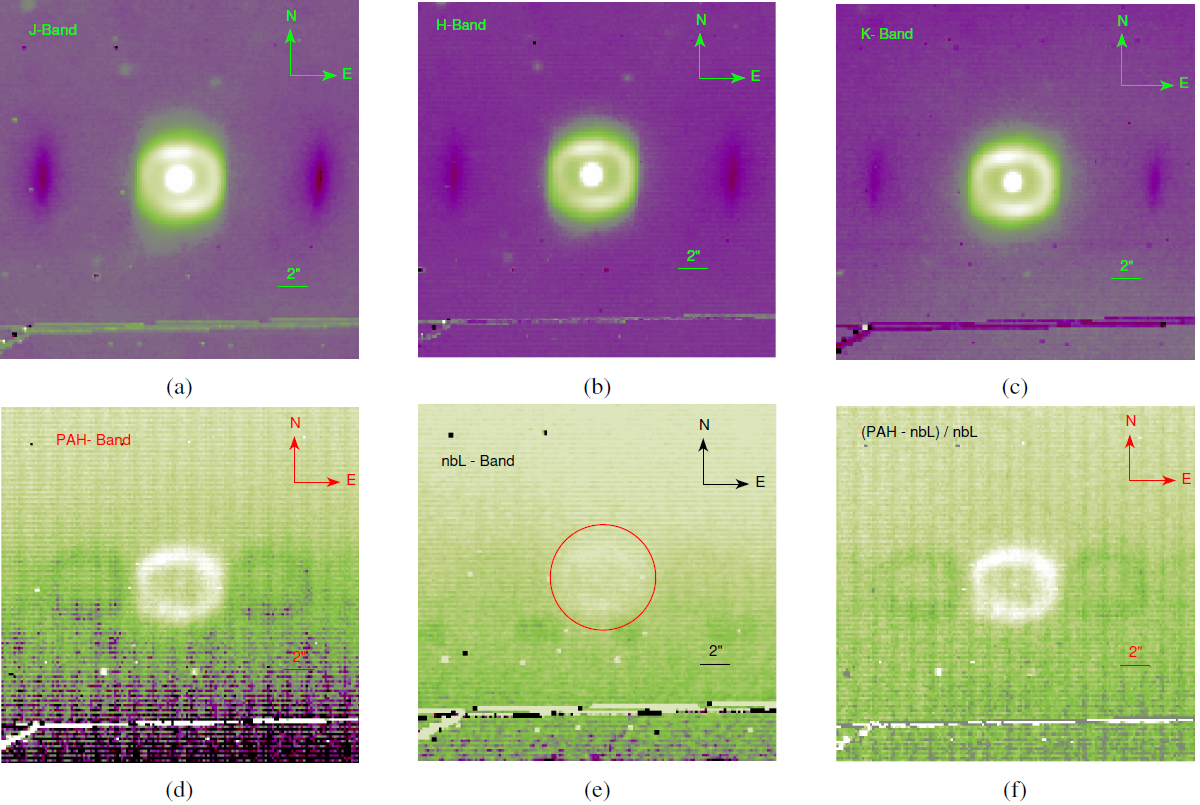}
        \caption{High-resolution images of BD +30$^\circ$~3639 taken with TIRCAM2 on 3.6m DOT: (a) J, 1.2 $\mu$m; (b) H, 1.65 $\mu$m (c) K, 2.19 $\mu$m; (d) PAH, 3.28 $\mu$m; (e) nbL, 3.59 $\mu$m; (f) normalized continuum difference image, $(I_{PAH} - I_{nbL})/I_{nbL}$.}
        \label{fig5}
\end{figure*}

The images in the five bands and the PAH normalized continuum difference image of AIB$_{3.28}$ emission are shown in Figure~\ref{fig5}. The bright central object and the surrounding relatively smooth nebula having similar morphology is seen in J, H and K filters. All the images have similar rectangular-ring shaped nebular emission feature and have two bright lobes in northern and southern parts. The northern lobe is slightly brighter than southern lobe in all observed bands \citep{Hora1993}. The deviation from spherical symmetry of the nebula is also suggested by the rectangular shapes in VLA observations \citep{Bryce1997} and in spatial distribution of emission in radio continuum \citep{Basart1987, Masson1989}.

The angular diameter of the BD $+$30$^\circ$~3639 is $\sim$8$^{\prime\prime}$ for J, H bands while the angular size in K, PAH and nbL bands is $\sim$6$^{\prime\prime}\hspace{-0.13cm}$.9 (see Table~\ref{tbl-5}). The central object brightness extends up to $\sim$2$^{\prime\prime}$ in the J, H, K band images. The cavity immediately surrounding the central star, measured from the bright exterior of the central object to the inner edge of the shell, is $\sim$0$^{\prime\prime}\hspace{-0.13cm}$.6 along N-S and is $\sim$1$^{\prime\prime}$ along the E-W directions (Table~\ref{tbl-5}). 

As the central star is bright in J, H, K bands, the pixel count ratio of the central object to that of the surrounding nebula is obtained. The central object is nearly 4 times brighter in J and H while in the K band it is only $\sim$1.7 times brighter (Table~\ref{tbl-5}). The angular thickness of the nebular ring is $\sim$1$^{\prime\prime}\hspace{-0.13cm}$.4.

\begin{figure*}
     \centering
         \includegraphics[width=\textwidth]{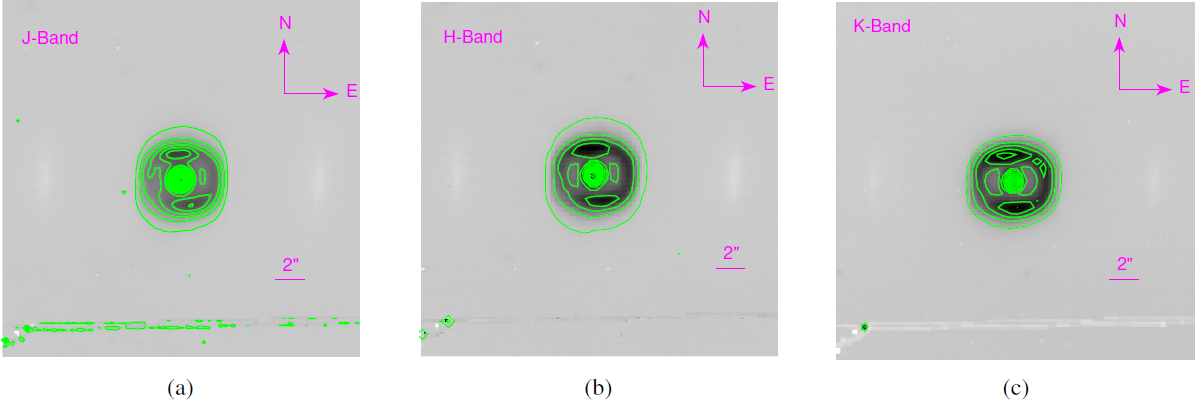}
  \caption{Contour images of BD +30$^\circ$~3639  at: (a) J, 1.2 $\mu$m; (b) H, 1.65 $\mu$m (c) K, 2.19 $\mu$m. Contours are separated by (a) 3.5\%, (b) 5\% and (c) 6\% of corresponding peak values.}
        \label{fig6}
\end{figure*}

The J, H, K contour images, shown in Figure~\ref{fig6}, point to similar morphology but the position of the intense lobe towards south shifts slightly westward as the wavelength increases. This may be indicative of different matter distribution and outflow.

\section{\textbf{Discussion}}

The young planetary nebulae NGC 7027 and BD $+$30$^\circ$~3639 are among the most studied and typical objects of their class. Both are in a similar stage of their evolution, with a carbon rich and molecular circumstellar medium. The first spectroscopic detection of the PAH related infrared emission bands was made on these PNe \citep{Gillett1973}. Both objects show similar 3.3~$\mu$m AIB feature, classified as class `A' \citep{Diedenhoven2004}. Despite these similarities the two objects are quite different in terms of their mass and temperature. While NGC 7027 is hot high excitation object, BD $+$30$^\circ$~3639 is of relatively lower mass and low excitation. Some basic data for the two objects are compared in Table~\ref{tbl-6}.

Both objects show cavity surrounding the central stars, where the high velocity winds from the hot core have pushed out the dust. As the winds from hot core move outward through the cavity, they are obstructed by torus of dust illuminating elliptical nebular shapes. The lower mass object BD $+$30$^\circ$~3639 shows a smoother shape compared to NGC 7027, which being of higher mass has evolved faster with complex morphology indicative of multiple outflow episodes \citep{Cox2002}.

\begin{figure*}
\centering
\begin{minipage}[b]{0.45\linewidth}
 \includegraphics[height=8cm, width=\columnwidth]{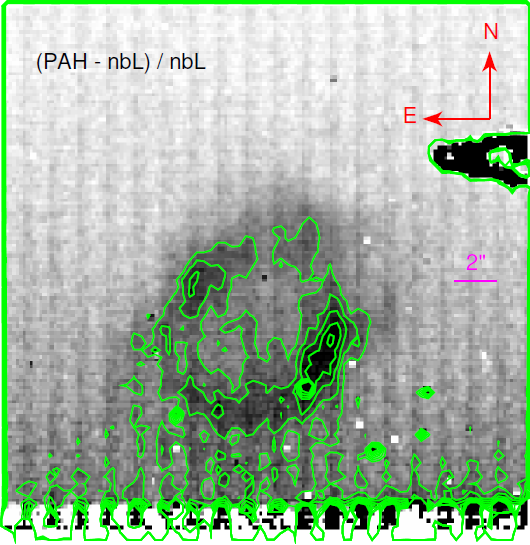}
 \caption{Contour image of NGC 7027 at 3.28$\mu$m. Contours are overlaid for clarity and are separated by 15\% of peak value.}
 \label{fig7}
\end{minipage}
\quad
\begin{minipage}[b]{0.45\linewidth}
 \includegraphics[height=8cm, width=\columnwidth]{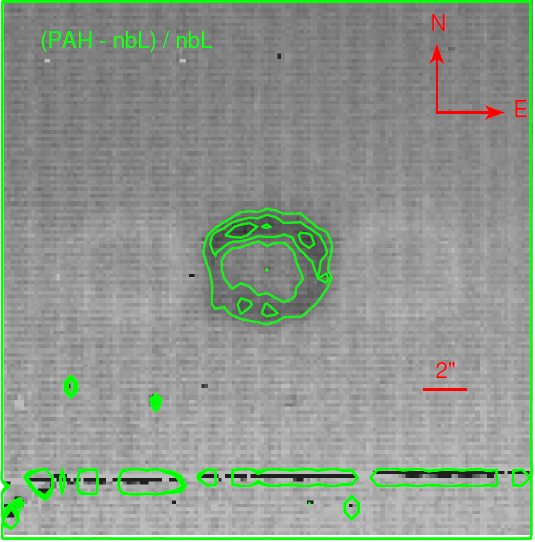}
 \caption{Contour image of  BD $+$30$^\circ$~3639 at 3.28$\mu$m. Contours are overlaid for clarity and are separated by 16\% of peak value.}
 \label{fig8}
\end{minipage}
\end{figure*}

Contour images of the AIB$_{3.28}$ feature in the two objects are shown in Figures~\ref{fig7} and \ref{fig8}. For NGC 7027 the two bright lobes are clearly seen in the NE and SW, that fall somewhat along the ellipse minor-axis \citep{Woodward1989}. This is consistent with the extended asymmetric emission in longer wavelength AIBs \citep{Lau2016}. In BD $+$30$^\circ$~3639 there are two bright regions in the northern part corresponding to the brighter JHK northern lobe. From the AIB$_{3.28}$ images average size of the PAH emission shell is obtained that defines the location of PAH molecules and also the region of possible PAH formation. The distance of this region from the central stars is about 6000 AU in NGC 7027 and about 5000 AU in BD $+$30$^\circ$~3639 (Table~\ref{tbl-6}).

The strength of the 3.3 $\mu$m AIB is much greater in NGC 7027 compared to that in BD $+$30$^\circ$~3639 \citep{Diedenhoven2004}. The circumstellar C/O ratio shows a linear correlation with the equivalent width of the feature \citep{Smith2008}, consistent with the higher C/O ratio in NGC 7027 (Table~\ref{tbl-6}). It is well understood that the AIBs result from a mixture of neutral and ionized PAHs \citep{Allamandola1999, Tielens2008, Rastogi2013}. Laboratory and computational studies show that the 3.3 $\mu$m feature is prominent only in neutral PAHs and cations have negligible intensity in this band \citep{Langhoff1996, Hudgins2002, Pathak2007, Maurya2015}. Therefore, the similar 3.3 $\mu$m AIB feature in the two objects indicate a similar population of predominantly neutral PAHs. Indeed, for BD $+$30$^\circ$~3639, a low-ionization fraction of PAHs is interpreted observationally \citep{Bernard1994, Matsumoto2008}, while an even distribution of PAH neutrals and cations is indicated for the hotter NGC 7027 \citep{Maragkoudakis2020}.

Though the 3.28 $\mu$m AIB in the two objects is of similar class `A', the 6.2 $\mu$m AIB profile is altered. In NGC 7027 the 6.2 $\mu$m feature is classified as `A' type, but in BD $+$30$^\circ$~3639 it is of `B' type \citep{Peeters2002, Diedenhoven2004}. As the features vary with PAH size and charge states \citep{Pathak2008, Maragkoudakis2020}, this difference in the 6.2 $\mu$m profile points towards variation in the PAH populations of the two objects. Generally PAH formation via bottom-up approach is envisaged in C-rich circumstellar shells of young PNe \citep{Frenklach1989, Cherchneff1992, Cernicharo2004} and in high excitation regions PAHs may also form by top-down grain fragmentation. In BD $+$30$^\circ$~3639 the C-rich stellar winds trigger PAH formation \citep{Matsumoto2008}, while in NGC 7027 rapid PAH formation is indicated from grain-grain collisions in the post-shock environment \citep{Lau2016}.

The route to formation of fresh PAHs in C-rich PNe \citep{Frenklach1989, Cherchneff1992} indicates intermediate compounds as PAHs with side groups. In such PAHs e.g. with vinyl or phenyl substitution \citep{Maurya2015, Rashmi2018} modes appear closer to 6.3~$\mu$m, that is similar to the class `B' 6.2~$\mu$m AIB. These PAH derivatives could be part of the PAH population in BD $+$30$^\circ$~3639. The PAH derivatives with aliphatic groups could also explain the step-like rise in continuum at around 3.6 $\mu$m by about 10\% of the 3.28 $\mu$m peak \citep{Ohsawa2016}. The higher excitation and advanced evolution in NGC 7027 lead to a mixed PAH population including more processed and ionized PAHs.

\section{Conclusion}

The observations of diffused PAH emission are difficult and challenging from ground-based telescopes. The PNe images are successfully obtained in J, H, K, PAH and nbL bands. The present observations are among the first detections in PAH band ($\lambda$\textsubscript{cen} $\sim$3.28 $\mu$m, $\bigtriangleup\lambda$ $\sim$0.06 $\mu$m) with the TIRCAM2 on DOT. The nbL band ($\lambda$\textsubscript{cen} $\sim$3.59 $\mu$m, $\bigtriangleup\lambda$ $\sim$0.07 $\mu$m) images show continuum emission features of the surrounding PNe medium. The final frames showing the PNe in these bands were obtained by combining all dithered frames. The observations confirm the presence of emission due to aromatic C-H stretch vibrations at 3.28 $\mu$m and also throw light on the morphology of the objects and location of PAHs.

The JHK observations in NGC 7027 reveal elliptical morphology and features extending up to 8$^{\prime\prime}$ from the central star. BD +30$^\circ$~3639 shows a rectangular-ring shaped nebula, with angular size of $\sim$8$^{\prime\prime}$extending in J and H filters. A smaller extent of $\sim$6$^{\prime\prime}\hspace{-0.13cm}$.9 is observed in K, PAH and nbL bands. The continuum normalized AIB$_{3.28}$ feature indicates that the location of PAHs is about 6000 AU and 5000 AU from the central star in NGC 7027 and BD $+$30$^\circ$~3639 respectively. Analysis of the 6.2~$\mu$m AIB feature variations and its classification suggests a dominantly neutral and freshly formed PAH population in BD $+$30$^\circ$~3639 while in NGC 7027 there is higher ionization and also possible PAH production by grain fragmentation.

\vspace{-1.5em}

\section*{Acknowledgements}
This work is done under MoU between ARIES and DDU Gorakhpur University. RA acknowledges financial support from University Grants Commission, New Delhi under the Rajiv Gandhi National Fellowship scheme.

\vspace{-0.5em}


\bibliographystyle{spbasic}      


\newpage

\clearpage

\begin{table*}
\renewcommand{\arraystretch}{1.3}
\centering
\tabularfont
\caption{TIRCAM2 Observation Details}\label{tbl-2}
\begin{tabular}{cccccc}
\topline
\textbf{Object Name}&\textbf{Filter}&\textbf{Dither Position}&\textbf{No. of frames}&\textbf{Exposure Time (sec)}&\textbf{Total Time Integration (sec)}\\
\midline
NGC 7027& J &3&4&20& 3 x 4 x 20 = 240\\
&H&3&4&20& 3 x 4 x 20 = 240\\
&K&3&7&12&3 x 7 x 12 = 252\\
&PAH&3&300&0.05&3 x 300 x 0.05 = 45\\
&nbL&3&300&0.05&3 x 300 x 0.05 = 45\\
\hline
BD $+$30$^\circ$~3639&J&3&3&8& 3 x 3 x 8 = 72\\
&H&3&3&8& 3 x 3 x 8 = 72\\
&K&3&5&5& 3 x 5 x 5 = 75\\
&PAH&3&300&0.05&3 x 300 x 0.05 = 45\\
&nbL&3&300&0.05&3 x 300 x 0.05 = 45\\
\hline
\end{tabular}
\end{table*}

\begin{table*}
\renewcommand{\arraystretch}{1.3}
\centering
\tabularfont
\caption{JHK Photometry of NGC 7027 and BD $+$30$^\circ$~3639}\label{tbl-3}
\begin{threeparttable}
\begin{tabular}{|p{2.2cm}|p{3.2cm}|p{2.0cm}|p{2.5cm}|p{2cm}|p{2cm}|}
\hline
\textbf{Objects}&\textbf{Standard Object}\tnote{*}&\textbf{Filter}&\textbf{Aperture Radius}&\textbf{Magnitude}&\textbf{Mag. Error}\\
&&&(arcsec)&&\\
\hline
NGC 7027&FS04 (TYC 30-78-1) & J &10.55&6.86&0.003 \\
&&H&8.44&6.77&0.004 \\
&&K\tnote{**}&-&-&-\\
\hline
BD $+$30$^\circ$~3639&FS25 (TYC 351-60-1) &J&4.14&9.72&0.005\\
&&H&4.23&9.44&0.012\\
&&K&3.43&8.21&0.009\\
\hline
\end{tabular}
\begin{tablenotes}\footnotesize
\item[*]\citet{Hunt1998}
\item[**]magnitude could not be obtained as a large part of the K-band image overlaps bad patches in the lower side of the array.
\end{tablenotes}
\end{threeparttable}
\end{table*}

\begin{table*}
\renewcommand{\arraystretch}{1.3}
\centering
\tabularfont
\caption{Size of NGC 7027}\label{tbl-4} 
\begin{threeparttable}
\begin{tabular}{ |m{1cm}|m{1.2cm}|m{1cm}|m{1cm}|m{1.5cm}|m{1.6cm}|m{1.5cm}|m{1.5cm}|m{1.2cm}|m{1.8cm}|}
\hline
Filter & Angular diameter&\multicolumn{2}{c|}{Elliptical radius} &\multicolumn{5}{c|}{Distance between central star and inner edge of the shell}&Extended asymmetric region from center\\ \cline{3-9}
&&Semi major axis& Semi minor axis & NE- Direction & NW- Direction& SE- Direction& SW-Direction&Average distance&\\
&(arcsec)&(arcsec)&(arcsec)&(arcsec)&(arcsec)&(arcsec)&(arcsec)&(arcsec)&(arcsec)\\
\hline
J & 14.03 & 6.76 & 4.12&1.69&2.84&2.80&2.61&2.49 &$\sim$ 8.2\\
H & 13.92 & 6.75 & 4.57&1.71&2.89&2.79&2.06&2.36&$\sim$8.2\\
K & 12.83 & 6.75 & 4.57&1.51&3.12&3.09&2.27&2.49&$\sim$8.0\\
AIB$_{3.28}$\tnote{*} & 14.03 & 7.14 & 5.19&1.79&3.05&2.81&3.11&2.69&$\sim$ 7.5\\
nbL & 14.03 & 7.14& 5.19&1.79&3.05&2.18&3.11&2.69&$\sim$ 7.5 \\ 
\hline
\end{tabular}
\begin{tablenotes}\footnotesize
\item[*]AIB$_{3.28}$ = $(I_{PAH} - I_{nbL})/I_{nbL}$
\end{tablenotes}
\end{threeparttable}
\end{table*}

\begin{table*}
\renewcommand{\arraystretch}{1.35}
\centering
\tabularfont
\caption{Size of BD $+$30$^\circ$~3639}\label{tbl-5} 
\begin{threeparttable}
\begin{tabular}{ |m{1.5cm}|m{1.1cm}|m{1.1cm}|m{2.5cm}|m{2.5cm}|m{2.5cm}|m{1.5cm}|m{1.5cm}|}
\hline
Filter & Angular Diameter&Angular size of central object &\multicolumn{3}{c|}{Distance between central star and inner edge of the shell}&Ratio of pixel values of the central star \& nebula&Thickness of ring-shaped emission\\ \cline{4-6}
&(arcsec) &(arcsec) &N-S Direction& E-W Direction&Average distance& & (arcsec)\\
\hline
J &8.28&2.20&0$^{\prime\prime}\hspace{-0.13cm}$.41&0$^{\prime\prime}\hspace{-0.13cm}$.82&0$^{\prime\prime}\hspace{-0.13cm}$.62&4.19 &1.32\\
H &8.45&2.10&0$^{\prime\prime}\hspace{-0.13cm}$.40&1$^{\prime\prime}\hspace{-0.13cm}$.06&0$^{\prime\prime}\hspace{-0.13cm}$.73&4.17 &1.33\\
K &6.86&1.98&0$^{\prime\prime}\hspace{-0.13cm}$.34&0$^{\prime\prime}\hspace{-0.13cm}$.83&0$^{\prime\prime}\hspace{-0.13cm}$.59&1.74 &1.36 \\
AIB$_{3.28}$\tnote{*} &6.93&-&1$^{\prime\prime}\hspace{-0.13cm}$.29&1$^{\prime\prime}\hspace{-0.13cm}$.70&1$^{\prime\prime}\hspace{-0.13cm}$.50&-&1.42\\
nbL &6.93&-&1$^{\prime\prime}\hspace{-0.13cm}$.29&1$^{\prime\prime}\hspace{-0.13cm}$.70&1$^{\prime\prime}\hspace{-0.13cm}$.50&-&1.42\\ 
\hline
\end{tabular}
\begin{tablenotes}\footnotesize
\item[*]AIB$_{3.28}$ = $(I_{PAH} - I_{nbL})/I_{nbL}$
\end{tablenotes}
\end{threeparttable}
\end{table*}

\begin{table*}
\renewcommand{\arraystretch}{1.5}
\centering
\tabularfont
\caption{Basic data of NGC 7027 and BD $+$30$^\circ$3639}\label{tbl-6} 
\begin{threeparttable}
\begin{tabular}{ |m{3.5cm}|m{3.5cm}|m{3.5cm}|}
\hline
Properties&NGC 7027 & BD $+$30$^\circ$3639\\
\hline
Temperature&200000 K \textsuperscript{(a)} &47000 K \textsuperscript{(d)}\\
Mass&3 - 4 \(M_\odot\) \textsuperscript{(b)}&$<$ 3 \(M_\odot\) \textsuperscript{(e)}\\
Distance&880 $\pm$ 150 pc \textsuperscript{(c)}&1520 $\pm$ 210 pc \textsuperscript{(f)}\\
Age of the Nebula&600 years \textsuperscript{(c)}&600-800 years \textsuperscript{(g)}\\
Shell expansion velocity&17 Km/s \textsuperscript{(c)}&22$\mp$4 Km/s \textsuperscript{(h)}\\
Nebula temperature&14500 K \textsuperscript{(c)}&8500 K \textsuperscript{(e)}\\
Electron density&7x10$^4$ cm$^3$ \textsuperscript{(c)} & 11x10$^3$cm$^3$ \textsuperscript{(e)}\\
C/O&1.46 \textsuperscript{(b)}&1.1 \textsuperscript{(i)}\\
Average location of AIB$_{3.28}$ emission from the central star&6170 AU&5270 AU\\
\hline
\end{tabular}
\begin{tablenotes}\footnotesize

\item{NGC 7027: \textsuperscript{(a)}\citet{Latter2000}, \textsuperscript{(b)}\citet{Salas2000}, \textsuperscript{(c)}\citet{Masson1989} } 
\item{{BD $+$30$^\circ$3639}:  \textsuperscript{(d)}\citet{Leuenhagen1996}, \textsuperscript{(e)}\citet{Salas2003}, \textsuperscript{(f)}\citet{Akras2012}, \textsuperscript{(g)}\citet{Jianyang2002}, \textsuperscript{(h)}\citet{Mellema2004},  \textsuperscript{(i)}\citet{Aller1995} }

\end{tablenotes}
\end{threeparttable}
\end{table*}

\end{document}